%% file: ckm_calderon.tex
\documentclass[12pt]{article}
\usepackage{epsfig}

\def\met{\ensuremath{\not\!\!{E_{T}}}}

\input econfmacros.tex
\def\Title#1{\begin{center} {\Large {\bf #1} } \end{center}}

\begin{document}

\Title{Top pair cross section measurements at the LHC}

\begin{center}
{\it Proceedings of CKM 2012, the 7th International Workshop on the CKM Unitarity Triangle, University of Cincinnati, USA, 28 September - 2 October 2012 
}
\end{center}

\bigskip


\begin{raggedright}  

{\it Alicia Calderon for the CMS and ATLAS collaborations. \index{Calderon, A.}\\
Instituto de Física de Cantabria (CSIC - UC) \\
University of Cantabria\\
39005 Santander, Spain}
\bigskip\bigskip
\end{raggedright}

\section{Introduction}

The most recent results on the measurements of ($t\bar{t}$) production and cross sections at 
7 TeV are presented. These are obtained using CMS~\cite{CMS} and ATLAS~\cite{ATLAS} data
collected in 2011. Recent results on the $t\bar{t}$ production cross section at 8 TeV using 2012 
CMS data are also presented. The $t\bar{t}$ inclusive cross sections are measured
in the lepton+jets, dilepton and fully hadronic channels, including the tau-dilepton
and tau+jets modes. The results are combined and confronted with precise theory
calculations.

\section{Dilepton and lepton+jets channels, with electrons and muons}

The dilepton final state has the lowest production rate, however a distinct event signature
allows for precise studies of the production rate. In both experiments, events are required to have
large $\met$ and at least two jets. In CMS, events are required to have at least one pair of oppositely signed
leptons (e, $\mu$) and at least one of the selected jets must be b-tagged. The cross section is extracted 
with a profile likelihood of the jet multiplicity, and the multiplicity of b-tagged jets. 
In ATLAS, selected events must either have exactly one pair of oppositely charged leptons, or one
lepton and one track-lepton candidate with opposite charge. The cross section is obtained with a 
profile likelihood technique combining eight channels, the dilepton and lepton+track channels where 
no jet is b-tagged, and the di-electron and di-muon channels with at least one b-tagged jet.

The $t\bar{t}$ cross section is measured as 161.9 $\pm$ 2.5(stat.) $^{+5.1}_{-5.0}$(syst.) $\pm$ 3.6(lumi.)pb in the CMS
analysis~\cite{CMS_Dilepton} with 2.3 $fb􀀀^{-1}$ of data and 176 $\pm$ 5(stat.) $^{+14}_{-11}$(syst.) $\pm$ 8(lumi.)pb 
in the ATLAS analysis~\cite{ATLAS_Dilepton} with 0.7 $fb􀀀^{-1}$ of data.

In addition, approximate NNLO QCD predictions for the cross section of top-quark 
pair production are used together with different parton distribution functions to extract the strong coupling constant, 
from the cross section measured by the CMS experiment~\cite{CMS_Dilepton_Alpha}. It probes the strong coupling 
with remarkable precision and in an energy regime that has been accessible to only a small number of measurements 
so far. The extracted value of is  0.1178 $^{+0.0046}_{-0.0040}$, in good agreement with the current world average. 

The balance of high production rate and clean event signature makes the lepton+jets final state
ideal for precision measurements. In both experiments, considered events are required to have
exactly one electron or muon and $\met$. In CMS the cross section is extracted with a 
simultaneous binned likelihood fit to the secondary vertex mass distribution for different jet multiplicities and 
number of b-tagged jets. In ATLAS, the cross section is extracted from a simultaneous likelihood fit to a likelihood
discriminant distribution for events with three, four and at least five jets.

In CMS~\cite{CMS_Lepton+Jets} the measured $t\bar{t}$ cross section is 164.4 $\pm$ 2.8(stat.) $\pm$ 11.9(syst.) $\pm$ 7.4(lumi.)pb
with 1.09 $fb􀀀^{-1}$ of data. The ATLAS~\cite{ATLAS_Lepton+Jets} analysis measures a $t\bar{t}$ cross section of
 179.0 $\pm$ 3.9(stat.) $\pm$ 9.0(syst.) $\pm$ 6.6(lumi.)pb with 0.70 $fb􀀀^{-1}$ of data.

\section{Dilepton and lepton+jets channels, with a hadronically decaying tau}

The following measurements consider the final state where the $\tau$ decays only hadronically. In both experiments, 
selected events are required to identify exactly one e or $\mu$, one $\tau$, large $\met$ and at least two jets, with one 
or more b-tagged jet. In CMS, the lepton and $\tau$ are required to have opposite sign charge. The cross section is
extracted with a counting method by subtracting the expected number of background events from
the number of observed data events. In ATLAS, the contribution from events with same sign (SS) lepton and $\tau$ is subtracted from
the opposite sign (OS) contribution, as background process are expected to have a similar amount
of events in both regions. The cross section is extracted with a $\chi^{2}$ fit to an OS-SS boosted decision
tree output distribution for events where the $\tau$ candidate has exactly one track and events where the
$\tau$ candidate has at least two tracks.

The CMS analysis~\cite{CMS_Dilepton_Tau} measures a $t\bar{t}$ cross section of 143 $\pm$ 14(stat.) $\pm$ 14(syst.) $\pm$ 3(lumi.)pb
with 2.2 $fb􀀀^{-1}$ of data, while the ATLAS analysis~\cite{ATLAS_Dilepton_Tau}  measures a  $t\bar{t}$ cross section of 
186 $\pm$ 13(stat.) $\pm$ 20(syst.) $\pm$ 7(lumi.)pb using 2.5 $fb􀀀^{-1}$ of data.

The $t\bar{t}$ production cross section is measured in the lepton+jets $\tau$ final state where the $\tau$ decays
hadronically. In CMS, selected events are required to have one identified $\tau$, $\met$ and at least four
jets, at least one of which is b-tagged. The cross section is extracted with a likelihood fit to a neural
network output distribution. In ATLAS, the $\tau$ is not explicitly identified. Instead selected events are required to have at
least five jets, at least two of which are b-tagged, and $\met$. One of the selected jets is chosen as the $\tau$
candidate. A distribution of the number of charged tracks associated to the $\tau$ candidate is then used
to separate true and misidentified $\tau$ candidates, as a $\tau$ decays preferentially to one or three charged
particles. A likelihood fit to this distribution is used to extract the cross section.

In CMS~\cite{CMS_Lepton+Jets_Tau}  a $t\bar{t}$ cross section of 156 $\pm$ 12(stat.) $\pm$ 33(syst.) $\pm$ 3(lumi.)pb is measured using
3.9 $fb􀀀^{-1}$ of data, while ATLAS~\cite{ATLAS_Lepton+Jets_Tau} measures 200 $\pm$ 19(stat.) $\pm$ 43(syst.)pb using 1.67 $fb^{-1}$ of data.

\section{Fully hadronic channel}

The challenge for studies of $t\bar{t}$ events in the fully hadronic final state is the separation of signal
from QCD multijet background events. In both experiments, events are required to have at least six
jets, at least two of which are identified as b-jets. The cross section is extracted with a unbinned likelihood fit to 
the reconstructed top quark mass, $m_t$. The top quark mass distribution is chosen to separate the $t\bar{t}$ 
and multijet contributions and is reconstructed from a kinematic fit.

The measured $t\bar{t}$ cross section is 136 $\pm$ 20(stat.) $\pm$ 40(syst.) $\pm$ 8(lumi.)pb in CMS~\cite{CMS_Hadronic}  
using 1.09 $fb􀀀^{-1}$ of data and 168 $\pm$ 12(stat.) $^{+60}_{-57}$(syst.) $\pm$ 7(lumi.)pb  in ATLAS~\cite{ATLAS_Hadronic} 
using 4.7 $fb􀀀^{-1}$ of data.

\section{Combined results for top pair production at $\sqrt s$ = 7 TeV}

In order to obtain a more precise estimate of the top quark pair production cross section, the
experiments combine the results of the measurements performed in each channel with a likelihood
fit. In both experiments, the fully hadronic, lepton+jets (e+jets, $\mu$+jets) and dilepton (ee,$\mu\mu$,e$\mu$)
results are combined. In CMS, the result in the dilepton $\mu\tau$ channel is also included, though the most
precise dilepton result is not yet included.

\begin{figure}[htb]
\begin{center}
\epsfig{file=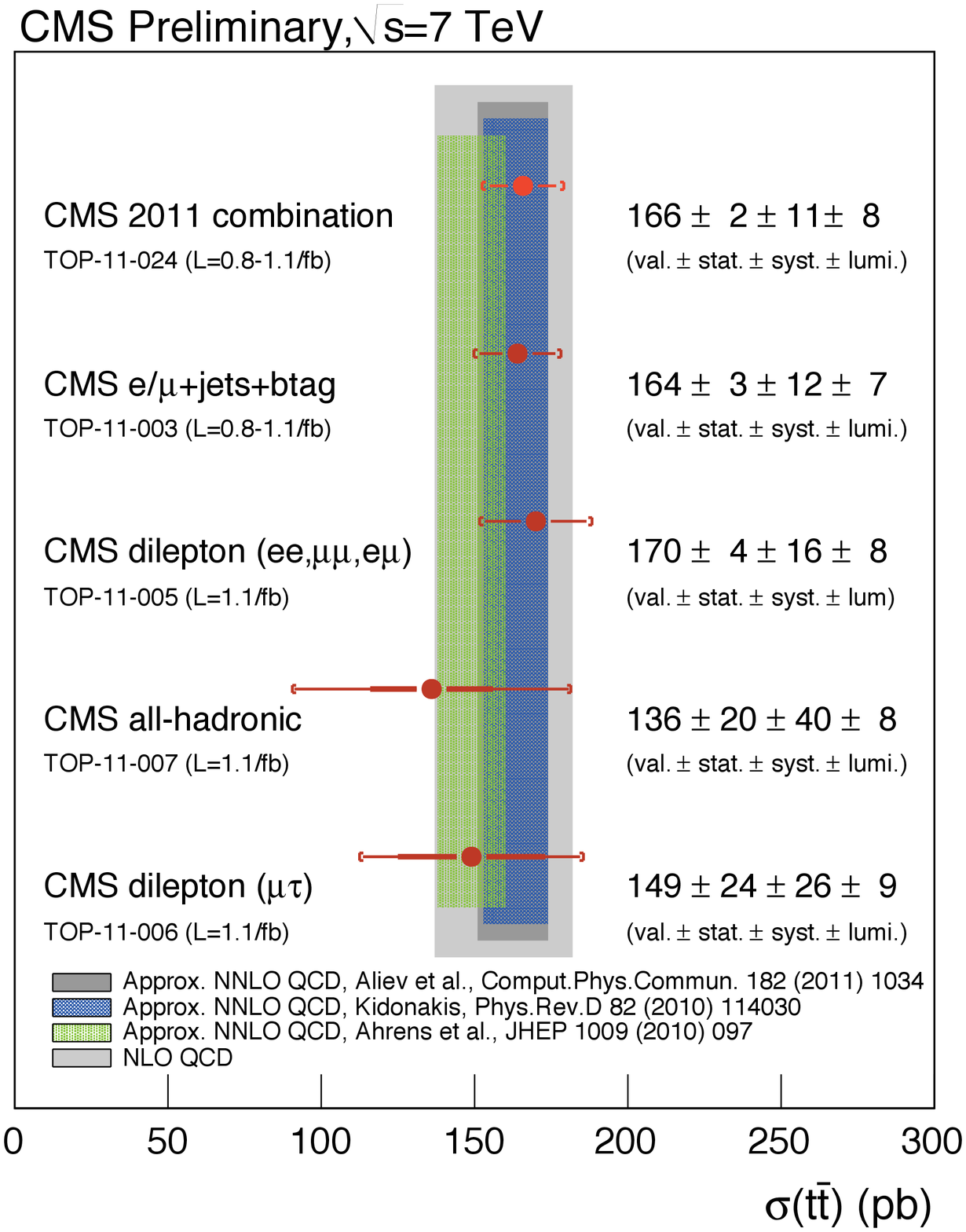,height=2.8in}
\epsfig{file=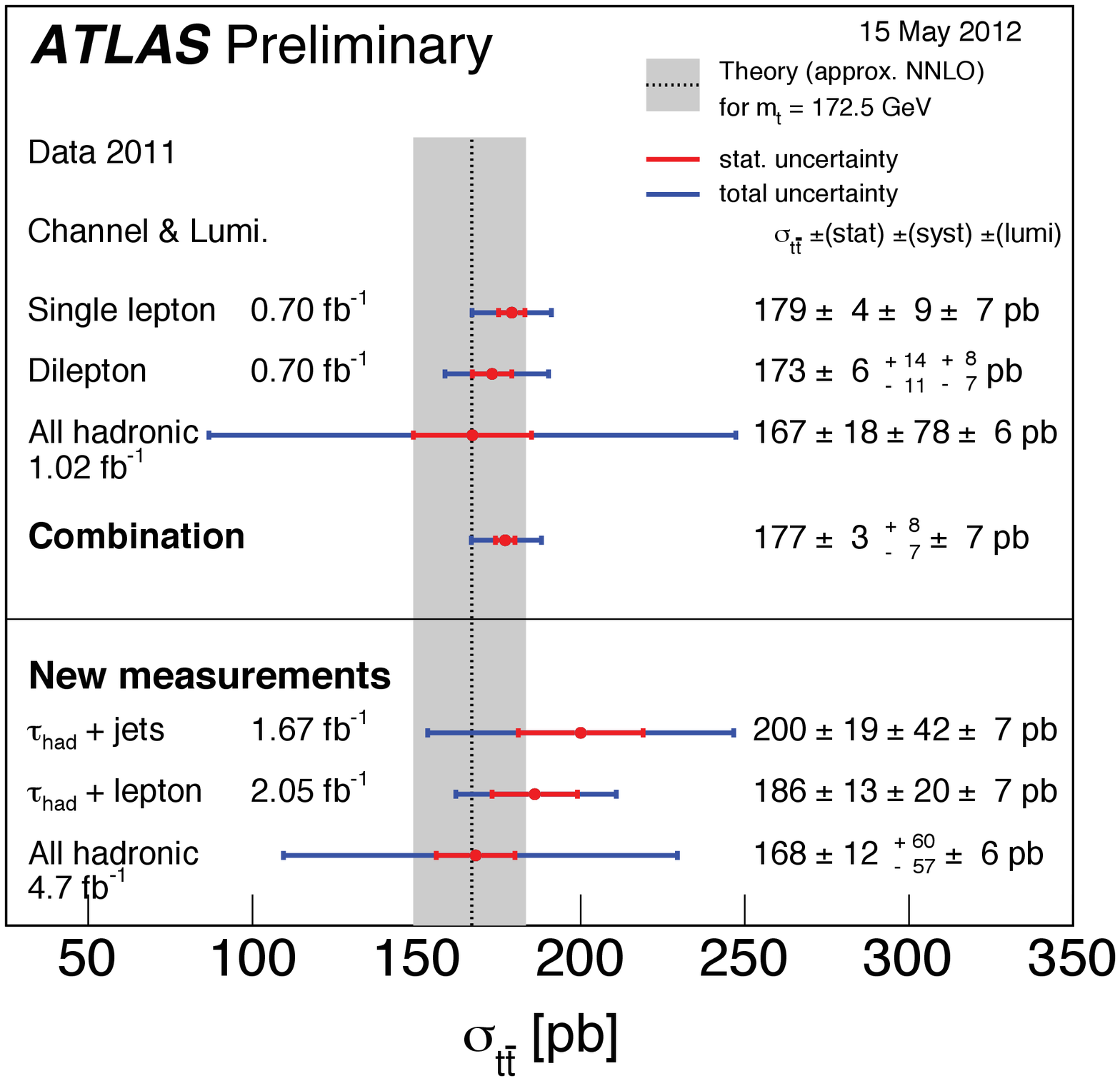,height=2.8in}
\caption{Results for measurements of the top quark pair production cross section performed in CMS (left)
and ATLAS (right) compared with Standard Model predictions}
\label{fig:Summary}
\end{center}
\end{figure}

The CMS combination~\cite{CMS_Combination} uses 0.8 to 1.1 $fb􀀀^{-1}$ of data and results in a $t\bar{t}$ cross section of
 165.8 $\pm$ 2.2(stat.) $\pm$ 10.6(syst.) $\pm$ 7.8(lumi.)pb . The ATLAS combination~\cite{ATLAS_Combination} is performed 
with 0.7 to 1.02 $fb􀀀^{-1}$ of data and results in a $t\bar{t}$
cross section of 177 $\pm$ 3(stat.) $^{+8}_{-7}$(syst.) $\pm$ 7(lumi.)pb .
The combined results, as well as the measurements in the individual channels, are compared
to the Standard Model prediction in Fig.~\ref{fig:Summary}. The Standard Model is found to be in agreement with
all measurements.

\section{CMS results for top pair production cross section at $\sqrt s$ = 8 TeV}

CMS experiment has measure the $t\bar{t}$ cross section in proton proton collisions at $\sqrt s$ = 8 TeV in 
the dilepton~\cite{CMS_8TeV_Dilepton} and in the lepton+jets~\cite{CMS_8TeV_Lepton+Jets} channels. The measurement 
in the dilepton channel is performed with events with two energetic leptons (electrons or muons) in the final state. Presence 
of the b-quark jets in the top-quark decays is required. The method used to extract the cross  section is a simple and 
robust counting experiment. The measured cross section is found to 
be 228 $\pm$ 9(stat.) $\pm$ 27(syst.) $\pm$ 10(lumi.)pb using 2.8 $fb^{-1}$ of data. In the lepton+jets channel the analysis 
is performed in the top quark pair decay channel with one isolated, high transverse momentum electron or muon, and at least 
four hadronic jets. At least one jet is required to originate from a b-quark. The measured cross section is extracted by a binned
likelihood fit of the invariant mass of the b-jet and the lepton system and it is found to 
be 227 $\pm$ 3(stat.) $\pm$ 10(syst.) $\pm$ 10(lumi.)pb using 2.41 $fb^{-1}$ of data.

\section{Conclusions}
A comprehensive examination of the top quark production processes has been performed by
the CMS and ATLAS collaborations with data collected at the LHC at $\sqrt s = 7$ TeV. Precision
measurements of $t\bar{t}$ production have been obtained in the dilepton and lepton+jets final states. 
First measurements of ttbar fully hadronic and tau final state have been performed. Initial CMS results at 
sqrt(s) = 8 TeV have been also presented. All measurements are found to be in agreement with Standard 
Model predictions.

\def\Discussion{
\setlength{\parskip}{0.3cm}\setlength{\parindent}{0.0cm}
     \bigskip\bigskip      {\Large {\bf Discussion}} \bigskip}
\def\speaker#1{{\bf #1:}\ }
\def\endDiscussion{}

\end{document}

%% file: econfmacros.tex
\def\beq{\begin{equation}}
\def\eeq#1{\label{#1}\end{equation}}
\def\eeqn{\end{equation}}

\def\beqa{\begin{eqnarray}}
\def\eeqa#1{\label{#1}\end{eqnarray}}
\def\eeqan{\end{eqnarray}}

\let\bar=\overbar

\def\Dslash{\not{\hbox{\kern-4pt $D$}}}
\def\dslash{\not{\hbox{\kern-2pt $\del$}}}

\def\msb{{\bar{\ssstyle M \kern -1pt S}}}

